\documentstyle[prd,eqsecnum,aps,epsf]{revtex}

\tightenlines

\newcommand{\be}{\begin{eqnarray}}
\newcommand{\beq}{\begin{eqnarray}}
\newcommand{\befg}{\begin{figure}}

\newcommand{\DSC}{D\hspace{-0.25cm}\slash_{\bot}}

\newcommand{\edfg}{\end{figure}}
\newcommand{\eeq}{\end{eqnarray}}
\newcommand{\ee}{\end{eqnarray}}

\newcommand{\QVBP}{\bar{Q}^{(+)}_{v^{\prime}} }
\newcommand{\QV}{Q^{(+)}_v}
\newcommand{\QVB}{\bar{Q}^{(+)}_v}

\begin{document}
\title{A More Precise Extraction of $|V_{cb}|$ in HQEFT of QCD}

\author{W. Y. Wang$^{* \dagger}$, 
Y. L. Wu$^{\dagger}$, Y. A. Yan$^\dagger$, M. Zhong$^\dagger$, 
Y. B. Zuo$^\dagger$}
\address{$*$ Department of Physics, Tsinghua University, Beijing 100084, China\\
$\dagger$ Institute of Theoretical Physics, Academia Sinica,
P.O. Box 2735, Beijing 100080, China}
\maketitle

\begin{abstract}

The more precise extraction for the CKM matrix element $|V_{cb}|$
in the heavy quark effective field theory (HQEFT) of QCD is
studied from both exclusive and inclusive semileptonic $B$ decays.
The values of relevant nonperturbative parameters up to order
$1/m^2_Q$ are estimated consistently in HQEFT of QCD. Using the
most recent experimental data for $B$ decay rates,  $|V_{cb}| $ is
updated to be $|V_{cb}| = 0.0395 \pm 0.0011_{\mbox{exp}} \pm
0.0019_{\mbox{th}} $ from $B\rightarrow D^{\ast} l \nu$ decay and
$|V_{cb}| = 0.0434 \pm 0.0041_{\mbox{exp}} \pm 0.0020_{\mbox{th}}
$ from $B\rightarrow D l \nu$ decay as well as $|V_{cb}| = 0.0394
\pm 0.0010_{\mbox{exp}} \pm 0.0014_{\mbox{th}} $ from inclusive
$B\rightarrow X_c l \nu$ decay.

\end{abstract}
\vspace{0.6cm} \hspace{1cm}{\bf PACS numbers}: 12.15.Hh, 12.39.Hg, 13.20.He

\hspace{0.9cm} {\bf Keywords:} $|V_{cb}|$, heavy quark effective field theory

\section{Introduction}

The Cabibbo-Kobayashi-Maskawa (CKM) matrix element $V_{cb}$ describes the rich phenomena of flavor-changing
transitions between the two heavy quarks $b$ and $c$. Its precise extraction has become a very important issue in
heavy flavor physics. Generally $|V_{cb}|$ is extracted by studying either exclusive or inclusive semileptonic $B$
decays. Since long distance contributions are involved in these decays, theoretical estimates of the
nonperturbative parameters are of crucial importance for a precise extraction of $|V_{cb}|$. Heavy quark symmetry
(HQS) \cite{symmetry,symmetry2,Is.Wi} and effective field theories of heavy quarks play a special role in such
estimates.

 In this short note,  we are going to provide a more precise extraction for $|V_{cb}|$ from both exclusive and inclusive
 semileptonic $B$ decays within the framework of HQEFT of QCD. This effective field theory can directly be derived
 from QCD\cite{YLW} by integrating out the small components but with carefully treating the quark and antiquark
 components. Recently, it has been shown that the HQEFT can actually be regarded as a large component QCD\cite{W9}.
 Of particular, this effective field theory has been put forward and successfully
 applied to various hadron processes \cite{W1,W2,W3,W4,W5,W6,W7,W8,bs1,bs2,mas}.
 The resulting effective Lagrangian and the heavy quark expansion (HQE) in the HQEFT of QCD
 appear to be different from the usual
 heavy quark effective theory (HQET) \cite{HG,HQ} in which the quark and antiquark
 components were dealt with separately, i.e., there is no quark-antiquark
 coupling terms considered in HQET. The application of the usual HQET in
 extracting $|V_{cb}|$ has been discussed by several groups and summarized in
 ref.\cite{PDG}. We shall further emphasize in this note
 the relevant new features of HQEFT with respect to the usual HQET . Our paper is organized as follows:
 $|V_{cb}|$ extraction from exclusive
and inclusive B decays will be discussed in section \ref{exclusive} and section \ref{inclusive}, respectively. The
previously obtained values for $|V_{cb}|$ will be updated by using the most recent experimental measurements. And
a brief conclusion will be presented in section \ref{conclusion}.

\section{$|V_{cb}|$ from exclusive decays}\label{exclusive}

The $B\to D^* (D)l\nu$ differential decay rates are
   \begin{eqnarray}
   \label{widthb2ds}
  \frac{d\Gamma(B\rightarrow D^{\ast}l\nu)}{d\omega} &=&\frac{G^2_F}{48\pi^3}(m_B-m_{D^{\ast}})^2
    m^3_{D^{\ast}}\sqrt{\omega^2-1}(\omega+1)^2 \nonumber\\
    &&  \times  [1+\frac{4\omega}{\omega+1}
    \frac{m^2_B-2\omega m_B m_{D^{\ast}}+m^2_{D^{\ast}}}
    {(m_B-m_{D^{\ast}})^2}]\vert V_{cb} \vert^2 {\cal F}^2(\omega) ,\\
\label{widthb2d}
    \frac{d\Gamma(B\rightarrow Dl\nu)}{d\omega}&=&\frac{G^2_F}{48\pi^3}(m_B+m_{D})^2
    m^3_{D}(\omega^2-1)^{3/2} \vert V_{cb} \vert^2 {\cal G}^2(\omega)
   \end{eqnarray}
   with
\begin{eqnarray}
   \label{zerorecoilF}
     {\cal F}(1) &=&
     \eta_{A} h_{A_1}(1) =\eta_{A} (1+\delta^*),\\
   \label{zerorecoilG}
     {\cal G}(1) &=& \eta_{V} [h_{+}(1)-\frac{m_B-m_D}{m_B+m_D}
      h_{-}(1)]
      =  \eta_{V} (1+\delta ) ,
\end{eqnarray}
where the QCD radiative corrections to two loops give the short distance coefficients $ \eta_A=0.960\pm 0.007 $
and $\eta_V=1.022\pm 0.004$ \cite{ac}. $\omega$ is the product of the four-velocities of the $B$ and $D^* (D) $
mesons, $\omega=v\cdot v'$. The weak transition form factors $h_{A_1}(\omega)$, $h_+(\omega)$ and $h_-(\omega)$
can be expanded in powers of $1/m_Q$ and represented by the heavy quark spin-flavor independent wave functions.
Then based on Eqs.(\ref{widthb2ds}) and (\ref{widthb2d}), $|V_{cb}|$ can be precisely extracted as long as the
form factors are reliably evaluated in some theoretical framework.

The HQE of the transition form factors are studied in detail up to order $1/m^2_Q$ in
HQEFT framework in Ref.\cite{W1}. When the contributions of operators containing two
gluon field strength tensors are omitted, at the zero recoil point $\omega=1$ the
relevant form factors can be written as
\begin{eqnarray}
\label{formfactorha1}
    h_{A_{1}}&=&1+\frac{1}{8\bar{\Lambda}^2}[\frac{1}{m_b}(\kappa_1+3\kappa_2)
     -\frac{1}{m_c}(\kappa_1-\kappa_2)]^2
     -\frac{1}{8m^2_b\bar{\Lambda}^2}
     (F_{1}+3F_{2}  \nonumber\\
&&     -2\bar{\Lambda}\varrho_{1}-6\bar{\Lambda}\varrho_{2})
     -\frac{1}{8m_c^2\bar{\Lambda}^2}(F_{1}-F_{2}-2\bar{\Lambda}\varrho_{1}
     +2\bar{\Lambda}\varrho_{2}) \nonumber\\
&&     +\frac{1}{4m_b m_c \bar{\Lambda}^2}(F_{1}+F_{2}-2\bar{\Lambda}\varrho_{1}
     -2\bar{\Lambda}\varrho_{2}) , \\
    \label{formfactorhz}
h_{+}&=& 1+\frac{1}{8\bar{\Lambda}^2}(\frac{1}{m_b}-\frac{1}{m_c})^2
     [(\kappa_1+3\kappa_2)^2-(F_1+3F_2) +2\bar{\Lambda}(\varrho_1+3\varrho_2)] , \\
\label{formfactorhf} h_{-}&=& 0,
    \end{eqnarray}
where the parameters in the rhs. of Eqs.(\ref{formfactorha1})-(\ref{formfactorhf}) are
defined in HQEFT \cite{W1}. For simplicity, when the variable $\omega$ is not written
explicitly, we refer to the zero recoil values of relevant functions, i.e.,
$h_{A_1}=h_{A_1}(1)$, $\kappa_1=\kappa_1(1)$, etc. The binding energy of a heavy meson
$M$ is defined as
\begin{equation}
\bar{\Lambda} \equiv \lim_{m_Q \to \infty} \bar{\Lambda}_M
  =\lim_{m_Q \to \infty} (m_M-m_Q).
\end{equation}

It is seen from Eqs.(\ref{formfactorha1}) and (\ref{formfactorhz})
 that automatically both the form factors $h_{A_1}$ and $h_+$ in HQEFT of QCD do not receive $1/m_Q$
order correction at the zero recoil point.  Another point to be emphasized is
Eq.(\ref{formfactorhf}). The vanished value of $h_-$ in HQEFT  arises from the fact
of partial cancellation between the $1/m_Q$ correction in the current expansion and
the $1/m_Q$ correction coming from the insertion of the effective Lagrangian into the
transition matrix elements \cite{W1}.  Such a cancellation is not observed in HQET
because in the latter framework the quark-antiquark couplings are not taken into
account explicitly.
Though the so-called Luke's theorem in HQET protects the weak transition
matrix elements from $1/m_Q$ order correction at zero recoil point, it does not
protect $h_-$ from such correction \cite{am}. As a result, in HQET framework, only the $B\to D^*
l\nu$ decay rate at zero recoil is strictly protected against $1/m_Q$ order correction
while $B\to Dl\nu$ decay rate is not. This is the main reason besides the experimental
considerations for the conclusion that the $B\to Dl\nu$ decay is not as favorable as
$B\to D^*l\nu$ decay for $|V_{cb}|$ extraction in HQET.

In HQEFT, Eq.(\ref{formfactorhf}) advocates reliable extraction of $|V_{cb}|$ from
both $B\to D^* l\nu$ and $B\to Dl\nu$ decays, because both rates of these decays do
not receive $1/m_Q$ order correction, as can be seen from
Eqs.(\ref{widthb2ds})-(\ref{formfactorhf}).

However, we note that the current world averages for $|V_{cb}|{\cal F}(1)$ and
 $|V_{cb}| {\cal G}(1)$ are \cite{PDG}
     \begin{eqnarray}
     \label{exp1}
     \vert V_{cb}\vert {\cal F}(1)&=&0.0383\pm 0.0005 \pm 0.0009 ,\\
     \label{exp2}
     \vert V_{cb}\vert {\cal G}(1)&=&0.0413\pm 0.0029 \pm 0.0027 ,
     \end{eqnarray}
so the current experimental data for $|V_{cb}| {\cal G}(1)$ receive larger
errors than those for $|V_{cb}|{\cal F}(1)$, which may lead to large
experimental uncertainty for the result of $|V_{cb}|$ extracted from $B\to Dl\nu$.

Now the value of $|V_{cb}|$ depends on our estimates of ${\cal F}(1)$ and ${\cal
G}(1)$. First of all, it has been shown \cite{W1} that some of the nonperturbative
parameters appearing in the rhs. of Eqs.(\ref{formfactorha1}) -(\ref{formfactorhf})
can be related to the heavy meson masses. Explicitly, we have
\begin{eqnarray}
\label{massB}
 \bar{\Lambda}_{D(B)}&=&
   \bar{\Lambda}-(\frac{1}{m_{c(b)}}
   -\frac{\bar{\Lambda}}{2m^2_{c(b)}})(\kappa_1+3\kappa_2)
  -\frac{1}{4m^2_{c(b)} \bar{\Lambda}}(F_{1}+3F_{2})
   +O(\frac{1}{m_{c(b)}^3}) ,  \\
\label{massBS}
\bar{\Lambda}_{D^{\ast}(B^{\ast})}&=&
   \bar{\Lambda}-(\frac{1}{m_{c(b)}}-\frac{\bar{\Lambda}}
  {2m^2_{c(b)}})(\kappa_1-\kappa_2)
   -\frac{1}{4m^2_{c(b)} \bar{\Lambda}}(F_{1}-F_{2})
   +O(\frac{1}{m_{c(b)}^3}) .
\end{eqnarray}
Thus some parameters can be determined from the heavy meson masses. A detailed study
has been presented in Ref.\cite{W1}. Here we would not repeat the analysis. It is
easily read from Ref.\cite{W1} that by using
\begin{eqnarray}
  m_b = 4.7 \mbox{GeV} , \;\;\; m_b-m_c = 3.36 \mbox{GeV} \;\;\;
 m_b+\bar{\Lambda}=5.21 GeV,
\end{eqnarray}
one obtains the following values for $\kappa_1$, $\kappa_2$ and $F_1$, $F_2$ :
\begin{eqnarray}
  \label{estfrommass}
  \kappa_1 \approx -0.615 \mbox{GeV$^2$}, \;\;\;
   \kappa_2 \approx 0.056 \mbox{GeV$^2$} ,\nonumber\\
   F_{1} \approx 0.917\mbox{{GeV}$^4$}  ,\; \;\;
   F_{2} \approx 0.004\mbox{{GeV}$^4$} ,
 \end{eqnarray}
 which agree well with the sum rule results: $\kappa_1=-0.5\pm 0.18 \mbox{GeV}^2$, $\kappa_2 \approx 0.08
\mbox{GeV}^2$ and $\bar{\Lambda}=0.53 \pm 0.08 \mbox{GeV}$ \cite{W3}.

Then up to order $1/m^2_Q$ only two parameters $\varrho_1$ and $\varrho_2$ remain
unknown. To have an estimation for them, we recall the definition of the
nonperturbative parameters. They are defined by the matrix elements as follows
\cite{W1}:
\begin{eqnarray}
\label{wavefunctiondef}
  <M^{\prime}_{v^{\prime}}\vert \QVBP\Gamma &&\frac{1}{iv\cdot D} (i\DSC)^2
      \QV \vert M_v> =-\kappa_1(\omega) \frac{1}{\bar{\Lambda}} Tr[\bar{\cal M}^{\prime}\Gamma {\cal M}]
     \nonumber\\
     && + \frac{1}{\bar{\Lambda}}Tr[\kappa_{\alpha\beta}(v,v^{\prime})
      \bar{\cal M}^{\prime}\Gamma P_{+}\frac{i}{2}\sigma^{\alpha\beta}{\cal M}] , \nonumber\\
  <M^{\prime}_{v^{\prime}}\vert \QVBP\Gamma &&\frac{1}{iv\cdot D}
     (i\DSC)(-iv\cdot D )(i\DSC)   \QV \vert M_v>
      =-\varrho_1(\omega)\frac{1}{\bar{\Lambda}}Tr[\bar{\cal M}^{\prime}\Gamma {\cal M}]
      \nonumber\\
    &&+\frac{1}{\bar{\Lambda}}Tr[\varrho_{\alpha\beta}(v,v^{\prime})\bar{\cal M}^{\prime}\Gamma P_{+}
      \frac{i}{2}\sigma^{\alpha\beta}{\cal M}] ,\nonumber \\
   <M^{\prime}_{v^{\prime}}\vert \QVBP\Gamma&& \frac{1}{iv\cdot D} (i\DSC)^2
     \frac{1}{iv\cdot D} (i\DSC)^2  \QV \vert M_v>
      =-\chi_1(\omega)\frac{1}{\bar{\Lambda}^2} Tr[\bar{\cal M}^{\prime}\Gamma {\cal M}]
     \nonumber\\
   &&  +\frac{1}{\bar{\Lambda}^2} Tr[\chi_{\alpha\beta}(v,v^{\prime})\bar{\cal M}^{\prime}\Gamma P_{+}
      \frac{i}{2}\sigma^{\alpha\beta}{\cal M}]
      + \cdots ,
  \end{eqnarray}
where the ellipsis in the last equation represents the
contributions of operators containing two gluon field
strength tensors.
The Lorentz tensor $\kappa_{\alpha\beta}$ is decomposed into scalar
factors as $\kappa_{\alpha\beta}(v,v')=i\kappa_2(\omega) \sigma_{\alpha\beta}
+\kappa_3(\omega)(v'_\alpha \gamma_\beta-v'_\beta \gamma_\alpha)$,
and $\varrho_{\alpha\beta}(v,v')$, $\chi_{\alpha\beta}(v,v')$
are decomposed similarly.
The functions $F_1$ and $F_2$ are defined as
\begin{eqnarray}
   \label{defineF1}
   F_{1}=\chi_{1}+2\bar{\Lambda}\varrho_{1} , \\
 \label{defineF2}
   F_{2}=\chi_{2}+2\bar{\Lambda}\varrho_{2} .
\end{eqnarray}

As will be further emphasized in the next section, the operator $ iv\cdot D $ in the matrix elements was
considered to give contribution of order the binding energy \cite{W1,W2}, i.e., $ iv\cdot D \sim v\cdot k \sim
\bar{\Lambda}$ \footnote{This relation is in the sense of the effective contributions of operators within some
matrix element.}, which has also been confirmed in the simple case by sum rule calculation \cite{W3}. Now if
taking this simple replacement, we get from Eqs.(\ref{wavefunctiondef})
\begin{eqnarray}
  \label{rho1estimate}
\varrho_1 \approx -\kappa_1 \cdot \bar{\Lambda} \approx
    0.3 \mbox{GeV}^3, \hspace{1.5cm}
\chi_1 \approx  \kappa^2_1 \approx 0.4 \mbox{GeV}^4 .
\end{eqnarray}
Note that our estimates in Eqs.(\ref{estfrommass}) and (\ref{rho1estimate}) are
consistent with the relation among them in Eq.(\ref{defineF1}).

In Eq.(\ref{estfrommass}) $F_2$ almost equals zero.
Of course one reason for this may be the possible partial
cancellation between $\chi_2$ and $2\bar{\Lambda} \varrho_2$
in Eq.(\ref{defineF2}).
But more importantly, we notice that $\chi_2$, $\varrho_2$
(and therefore $F_2$) are parameters characterizing the
chromomagnetic type operators, and the contributions of
these chromomagnetic operators are generally believed
to be small.

If using the same method of estimation as that for $\varrho_1$ and $\chi_1$, one gets
\begin{eqnarray}
\label{rho2estimate}
  \varrho_2 & \sim & -\kappa_2 \cdot \bar{\Lambda} \approx -0.03 \mbox{GeV}^3,\\
\label{chi2estimate}
 \chi_2  & \sim & \kappa^2_2 \approx 0.003  \mbox{GeV}^4.
\end{eqnarray}
It can be seen from the above equations that the two terms in Eq.(\ref{defineF2}) do
have partial cancellation. However, Eqs.(\ref{rho2estimate}), (\ref{chi2estimate}) and
(\ref{defineF2}) can not hold simultaneously. This indicates that
Eqs.(\ref{rho2estimate}) and (\ref{chi2estimate}) are only rough
estimates for the
chromomagnetic type parameters since these parameters are very small in magnitude.
Nevertheless, as $\varrho_2$ must be small and can not influence the final result for
$|V_{cb}|$ as significantly as $\varrho_1$ does, here we would first take
Eq.(\ref{rho2estimate}) in extracting $|V_{cb}|$ but leave more rigid determination of
$\varrho_2$ for future work.

Fig.1 presents $\delta^*$ as a function of $m_b+\bar{\Lambda}$ at the fixed quark mass
difference $m_b-m_c=3.36$GeV. The curves are relatively flat in the region around
$m_b+\bar{\Lambda}=5.2$GeV, which indicates a reliable extraction of $\delta^*$ around
this point. This value of $m_b+\bar{\Lambda}$ is in agreement with that in
Eq.(\ref{estfrommass}). Varying $m_b$ and $m_b-m_c$ in the ranges $4.6\mbox{GeV} \leq
m_b \leq 4.8\mbox{GeV}$ and $3.32\mbox{GeV} \leq m_b-m_c \leq 3.41 \mbox{GeV}$, and
taking $\varrho_1 \approx 0.3 \mbox{GeV}^3$ and  $\varrho_2 \approx -0.03
\mbox{GeV}^3$, we get
\begin{equation}
 \label{deltas1}
 \delta^{\ast}=0.01 \pm 0.04.
\end{equation}
From Eqs.(\ref{exp1}) and (\ref{deltas1}), we then obtain
\begin{equation}
  \label{vcbbds}
  \vert V_{cb} \vert = 0.0395 \pm 0.0011_{\mbox{exp}}
     \pm 0.0019_{\mbox{th}} .
\end{equation}

For $B\to Dl\nu$ decay, analogously, we obtain
\begin{eqnarray}
\delta=-0.07 \pm 0.04
\end{eqnarray}
 and
\begin{equation}
 \label{vcbbd}
   | V_{cb} | = 0.0434 \pm 0.0041_{\mbox{exp}} \pm 0.0020_{\mbox{th}} .
\end{equation}

Comparing Eqs.(\ref{vcbbds}) and (\ref{vcbbd}), we see that the results of $|V_{cb}|$
extracted from $B\to D^* (D)l\nu$ decays have similar theoretical errors. But large
difference exists between the central values of our present results,
Eqs.(\ref{vcbbds}) and (\ref{vcbbd}). However, due to the large experimental error
for $B\to Dl\nu$ channel, the $|V_{cb}|$ values extracted above are compatible with
each other. A better determination of $|V_{cb}|$ from $B\to Dl\nu$ decay is expected
when a more precise result for the quantity $|V_{cb}| {\cal G}(1)$ is obtained.

Note that in Ref.\cite{W1} $\varrho_1$ and $\varrho_2$ are not
estimated as above. There $\varrho_2 \approx 0.1 \mbox{GeV}^3$ is
assumed so that the central values for $|V_{cb}|$ extracted from
$B\to D^*(D)l\nu$ decays can be close to each other. Indeed,
$\varrho_2 \approx 0.1 \mbox{GeV}^3$ together with the
experimental averages (\ref{exp1}), (\ref{exp2}) lead to
$|V_{cb}|=0.0416 \pm 0.0011_{\mbox{exp}} \pm 0.0016_{\mbox{th}}$
from $B\to D^* l\nu$ decay and $|V_{cb}|=0.0417\pm
0.0040_{\mbox{exp}} \pm 0.0019_{\mbox{th}}$ from $B\to Dl\nu$
decay.

\section{$|V_{cb}|$ from inclusive decays}\label{inclusive}

Inclusive semileptonic $B$ decays is the other alternative to determine $|V_{cb}|$. In
the usual HQET the light quark in a hadron is generally treated as a spectator, which
does not affect the heavy hadron properties to a large extent. However, this treatment
may be one possible reason for the failure of HQET in some applications. For example,
the world average value for bottom hadron lifetime ratio $\tau(\Lambda_b)/\tau(B^0)$
can not be explained well in the usual framework of HQET \cite{mc,mjch}.

The dynamics of inclusive $B$ decays is also analyzed in detail in Refs.\cite{W2,W4}
within the HQEFT framework. Instead of simply applying the equation of motion for
infinitely heavy free quark, $iv\cdot D \QV=0$, we treat the heavy quark in a hadron
as a dressed particle, which means that the residual momentum $k$ of the heavy quark
within a hadron is considered to comprise contributions from the light degrees of
freedom. This simple picture is adopted to conveniently take into account the effects
of light degrees of freedom and the binding effects of heavy and light components of
the hadron but not deal with the complex dynamics of hadronization directly. As the
light degrees of freedom within the heavy hadron is relativistic, one has
\begin{equation}
k^0 \sim | {\bf k} |.
\end{equation}
 Explicitly, the momentum of a heavy hadron $H$ is represented as $P_H=m_Q v+k+k'$ with $k'$ being the momentum depending on
the heavy flavor and suppressed by the inverse of the heavy quark mass. This directly leads to the relation
between $v\cdot k$ and the binding energy,
\begin{eqnarray}
\bar{\Lambda}= \lim_{m_Q \to \infty} \bar{\Lambda}_H
  =\lim_{m_Q\to \infty} (m_H-m_Q)=v\cdot k ,
\end{eqnarray}
or
\begin{equation}
\langle iv\cdot D \rangle \equiv \frac{ \langle H_v|\QVB iv\cdot D
\QV|H_v\rangle }{2\bar{\Lambda}_H} \approx \bar{\Lambda} \ne 0.
\end{equation}

To acquire a good convergence of HQE, we perform the expansion
in terms of $k-v \langle iv\cdot D \rangle$ (or say, equivalently,
in terms of $1/(m_Q+\bar{\Lambda}))$.
Then the $B\to X_cl\nu$ decay rate is found to be \cite{W2,W4}
\begin{eqnarray}
\label{incwidth}
 \Gamma(B\to X_c l \nu)=\frac{G^2_F \hat{m}^5_b V^2_{cb}}{192\pi^3}
   \eta_{cl}(\rho,\rho_l,\mu) \{ I_0(\rho,\rho_l,\hat{\rho})+I_1(\rho,\rho_l,\hat{\rho})
\frac{\kappa_1}{3 \hat{m}^2_b}
-I_2(\rho,\rho_l,\hat{\rho})\frac{\kappa_2}{\hat{m}^2_b} \},
\end{eqnarray}
where $\hat{m}_b=m_b+\bar{\Lambda}$, $I_0$, $I_1$ and $I_2$ are functions of the mass
square ratios $\rho={m^2_c}/{\hat{m}^2_b}$, $\hat{\rho}^2={\hat{m}^2_c}/{\hat{m}^2_b}$
and $\rho^2_l={m^2_l}/{\hat{m}^2_b}$. Here the calculation is performed up to
nonperturbative order $1/\hat{m}^2_Q$ and perturbative order $\alpha^2_s$. The
function $\eta_{cl}$ characterizes QCD radiative corrections. Its two-loop results were obtained in Refs.\cite{mmm,mmmb}. $\hat{m}_b$ and $\hat{m}_c$ can be determined from
the meson masses via Eq.(\ref{massB}). $\kappa_2$ is often extracted from the known
$B-B^*$ mass splitting
\begin{equation}
\kappa_2 \simeq \frac{1}{8} (m^2_{B^{*0}}-m^2_{B^0} )\approx 0.06
\mbox{GeV}^2,
\end{equation}
which is consistent with Eq.(\ref{estfrommass}) and the sum rule
result \cite{W3}.

There are several points to be mentioned for Eq.(\ref{incwidth}). Firstly, in deriving
Eq.(\ref{incwidth}) the effects of light degrees of freedom are explicitly accounted
for in the picture of a dressed heavy quark in a hadron. Secondly, it is seen that the
next leading order contributions vanish in our HQE in terms of the inverse dressed
heavy quark mass, $1/\hat{m}_b$. Furthermore, our HQE in terms of $k-v \langle iv\cdot
D \rangle$ (or $1/\hat{m}_b$) has a good convergence. It is found that the
$1/\hat{m}^2_b$ order contributions induce only $-0.7\sim 5\%$ corrections to the
total width $\Gamma(H_b \to X_c e \bar{\nu})$. Therefore we conclude that the higher
order nonperturbative corrections can be safely neglected. Finally, now one needs only
to treat the dressed quark mass $\hat{m}_b=m_b+\bar{\Lambda}$ instead of considering the
uncertainties arising from the two quantities $m_b$ and $\bar{\Lambda}$ separately. Note
that these are the features of HQE in HQEFT. They can not be observed in the HQE in
the usual HQET, where one assumes $\langle iv\cdot D \rangle $ to be zero or of higher
order of $1/m_b$. In the latter case the next leading order corrections can be absent
only when the HQE is performed in terms of $1/m_b$, and the heavy quark mass $m_b$ and
the binding energy $\bar{\Lambda}$ have to be treated separately. One result of this
is that in HQET the theoretical prediction of the total decay width strongly depends
on the value of bottom quark mass $m_b$ and may have larger uncertainties than in
HQEFT.

Using Eq.(\ref{incwidth}), $|V_{cb}|$ can be extracted from
experimental data for inclusive decay rates. Fig.3 shows the
obtained values of
$\delta_{in} \equiv \Gamma(B\to X_c l\nu)/(|V_{cb}|^2) \times
10^{11}$ as a function of the energy scale
$\mu$ and the parameters $m_c$, $\kappa_1$. It is seen that the
extracted value of $|V_{cb}|$ depends on the energy scale $\mu$
weakly. So the main uncertainties come from $m_c$ and $\kappa_1$.
The curves in Fig.3(b) and Fig.3(c) have minimal value points, around
which the values of $\delta_{in}$ are favorable because they are
less sensitive to the variation of $m_c$ and $\kappa_1$.
Varying $m_c$ and $\kappa_1$ in the regions
$1.45\mbox{GeV} < m_c < 1.85\mbox{GeV}$, $-0.8\mbox{GeV}^2<\kappa_1
 < -0.4 \mbox{GeV}^2$, these favorable values of $\delta_{in}$
change in the range:
\begin{equation}
\label{deltainclusive} \delta_{in}=2.88 \pm 0.20 \mbox{ps}^{-1},
\end{equation}
where the central value is obtained at $m_c=1.65 \mbox{GeV}$ and
$\kappa_1=-0.6\mbox{GeV}^2$.

Then the $B^0$ lifetime $\tau(B^0)=1.540\pm 0.014 \mbox{ps}$ and
the most recent CLEO data for $B\to X_c e\nu$ branching ratio
$\mbox{Br}(B\to X_c e \nu)=(10.49\pm 0.17 \pm 0.43 ) \% $
\cite{bCLEO} yield
\begin{eqnarray}
|V_{cb}|=0.0394 \pm 0.0010_{\mbox{exp}} \pm 0.0014_{\mbox{th}}.
\end{eqnarray}

$|V_{cb}|$ extracted using different values of branching ratios is presented in Fig.4.
As mentioned above, the values at $m_c \approx 1.65$GeV should be favorable because in
this region the curves become less sensitive to the mass $m_c$. Interestingly, as
shown in Fig.3(c), when choosing $m_c=1.65$GeV we find that the curve of
$\delta_{in}$(or the resultant $|V_{cb}|$) as a function of $-\kappa_1$ reaches its
minimal (or maximal) point at $\kappa_1 \approx -0.6 \mbox{GeV}^2$. This value for $\kappa_1$ is
again in good agreement with that we used in section \ref{exclusive} and with that
obtained from sum rule calculation \cite{W3}.

Note that the mass $m_c$ discussed here arises from the charm
quark propagator in the HQE of the matrix elements. It should be the
pole mass of charm quark. Thus it is not surprising that the value
of $m_c$ obtained here is larger than the value for constituent
mass quoted in section \ref{exclusive}.

\section{Conclusion}\label{conclusion}

 We have presented a more precise extraction of $|V_{cb}|$ by studying the
exclusive and inclusive semileptonic $B$ decays up to the order of $1/m^2_Q$ in HQEFT of QCD. The nonperturbative
parameters in HQEFT up to the same order are estimated consistently from various considerations.

 It has been shown that in HQEFT of QCD $|V_{cb}|$ can be reliably extracted from exclusive
 decays $B\to D^* (D)l\nu$ with similar theoretical uncertainties, because neither of their
differential decay rates receives $1/m_Q$ order corrections at zero recoil point. In
studying inclusive $B$ decays, we treat the heavy quark in a hadron as a dressed
particle whose residual momentum comprises some effects from the light degrees of
freedom. This enables us to consider the effects of light components in the hadron but
still has a simple physical picture in application. In this framework, the $B\to X_c
l\nu$ total decay rate is protected from $1/\hat{m}_b$ order correction, and the
$1/\hat{m}^2_b$ order correction is very small. Furthermore, $m_b$ and $\bar{\Lambda}$
only appear in the form of dressed quark mass: $\hat{m}_b=m_b+\bar{\Lambda}$, which
must also reduce the uncertainties in our calculations as
 $\hat{m}_b = m_H [1 + O(1/m_Q^2) ]$.

Using the most recent experimental data for the B-meson decay rates, we have arrived
at the determination for $|V_{cb}|$ with
\begin{eqnarray*}
& & |V_{cb}|=0.0395 \pm 0.0011_{\mbox{exp}}\pm 0.0019_{\mbox{th}} \qquad \mbox{from} \quad  B\to D^*l\nu ,\\
& & |V_{cb}|=0.0434 \pm 0.0041_{\mbox{exp}}\pm 0.0020_{\mbox{th}} \qquad \mbox{from} \quad B\to Dl\nu ,\\
& &|V_{cb}|=0.0394 \pm 0.0010_{\mbox{exp}}\pm 0.0014_{\mbox{th}}
\qquad \mbox{from} \quad B\to X_c l\nu,
\end{eqnarray*}
where the result extracted from $B\to D l\nu$ decay receives a larger experimental
uncertainty than that from $B\to D^* l\nu $ decay but a similar theoretical
uncertainty as the latter. The result obtained from $B\to D^*l\nu$ decay agrees quite
well with that from inclusive $B\to X_c l\nu$ decay.

These results then give the average
\begin{equation}
 \label{vcbaverage}
 |V_{cb}|=0.0402 \pm 0.0014_{\mbox{exp}} \pm 0.0017_{\mbox{th}}.
\end{equation}
Alternatively it can be represented as
\begin{equation}
 A=0.83 \pm 0.07
\end{equation}
in the Wolfenstein parameterization $|V_{cb}|=A \lambda^2$ with
$\lambda=|V_{us}|=0.22$.
\\
\\
\\


\centerline{\bf Acknowledgement}

This work was supported in part by the BEPC National Lab Opening
Project, the key projects
of Chinese Academy of Sciences and National Science Foundation of China
(NSFC).


\newpage
\mbox{}
\begin{figure}
\vspace{5cm}
\epsfxsize=7.0cm
\epsfysize=4.0cm
\centerline{
\epsffile{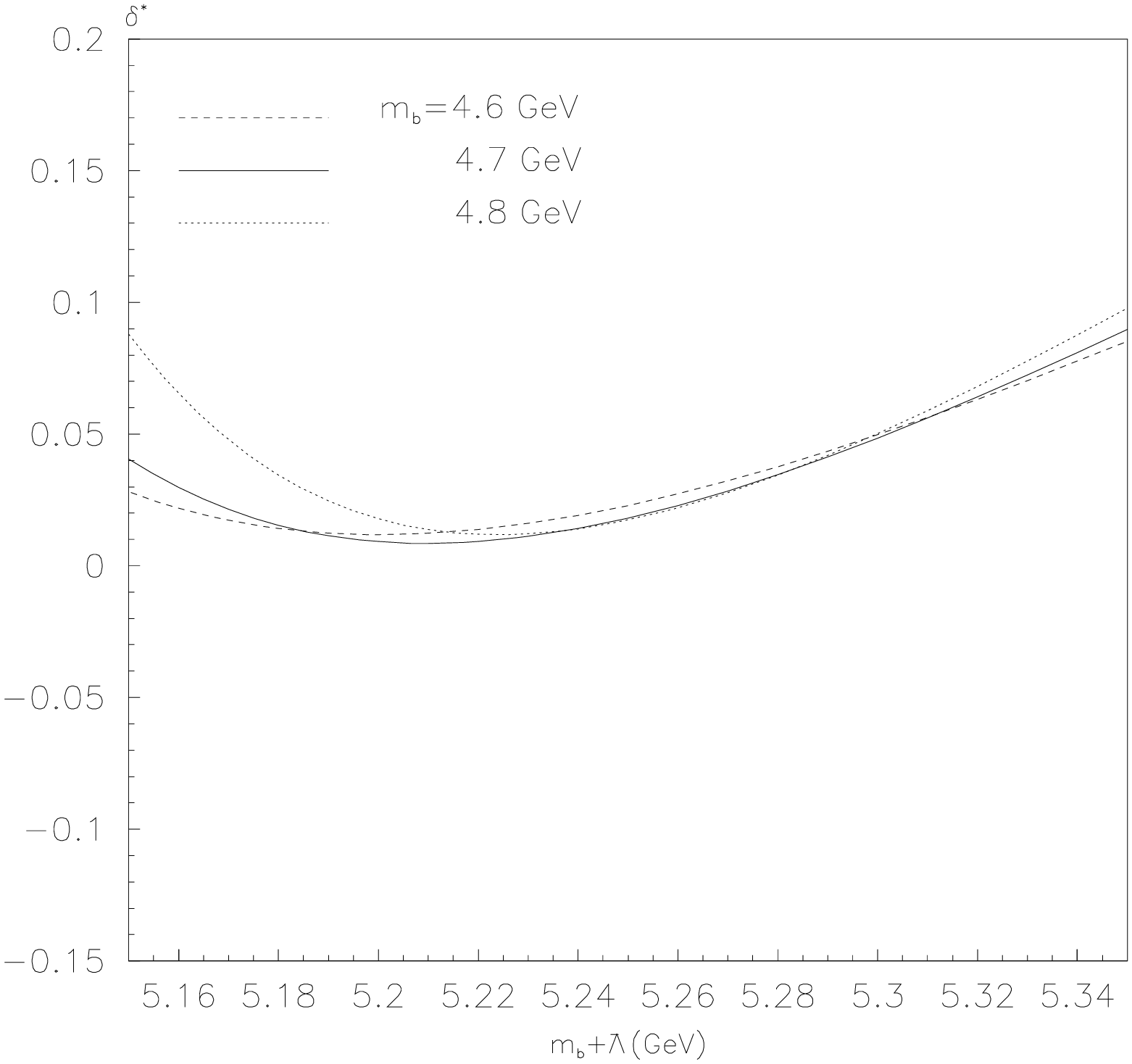} }
\vspace{0cm}
\caption{
  $\delta^{\ast}$ as a function of $m_b+\bar{\Lambda}$ at
  $m_b-m_c=3.36\mbox{GeV}$.
}
\end{figure}

\newpage
\mbox{}
\begin{figure}
\vspace{5cm}
\epsfxsize=13.0cm
\epsfysize=12.0cm
\centerline{
\epsffile{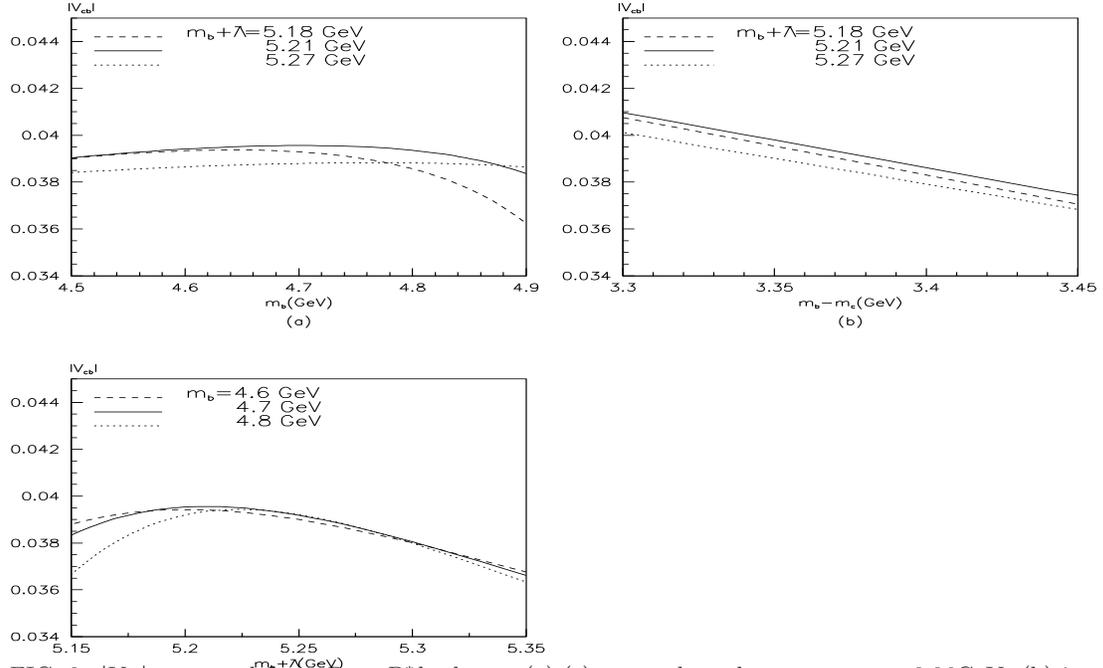}}
\vspace{-2cm}
\caption{
  $| V_{cb} |$ extracted from $B\rightarrow D^{\ast}l\nu$
  decay.
  (a),(c) are evaluated at $m_b-m_c=3.36\mbox{GeV}$; (b) is
  evaluated at $m_b=4.7\mbox{GeV}$.
}
\end{figure}

\newpage
\mbox{}
\begin{figure}
\vspace{5cm}
\epsfxsize=13.0cm
 \epsfysize=12.0cm
 \centerline{
 \epsffile{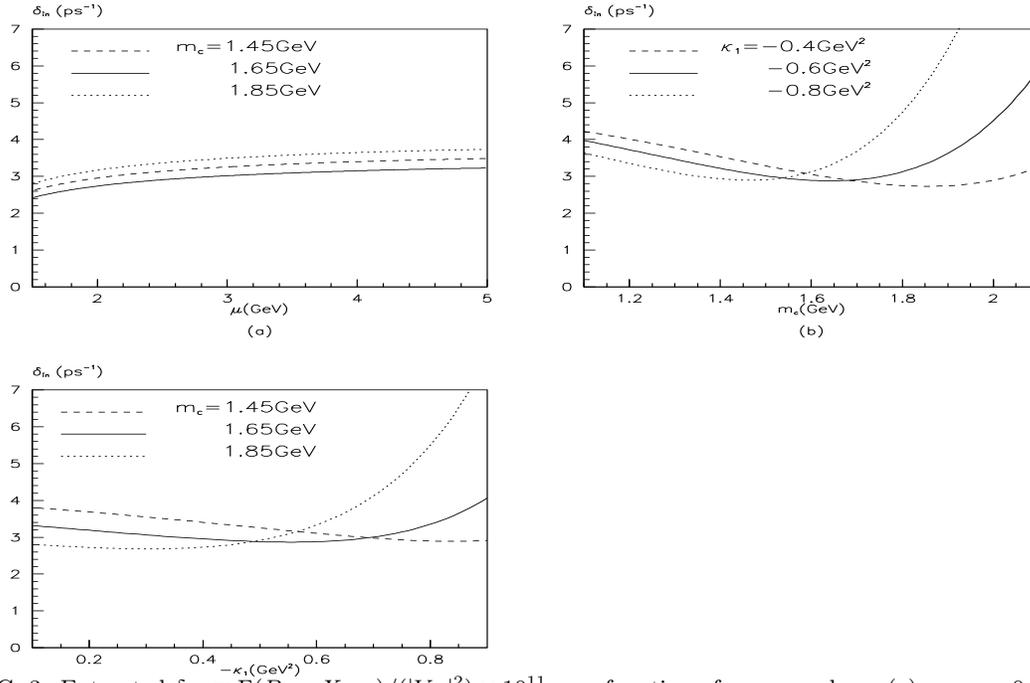}}
\vspace{-2cm}
 \caption{
  Extracted $\delta_{in}\equiv \Gamma(B\to X_c e \nu)/(|V_{cb}|^2)
\times 10^{11} $ as a function of $\mu$, $m_c$ and $\kappa_1$.
   (a): $\kappa_1=-0.6\mbox{GeV}^2$;
   (b) and (c): $\mu=2.4$GeV.
  }
\end{figure}

\newpage
\mbox{}
\begin{figure}
\vspace{5cm}
\epsfxsize=7.0cm
\epsfysize=4.0cm
\centerline{
\epsffile{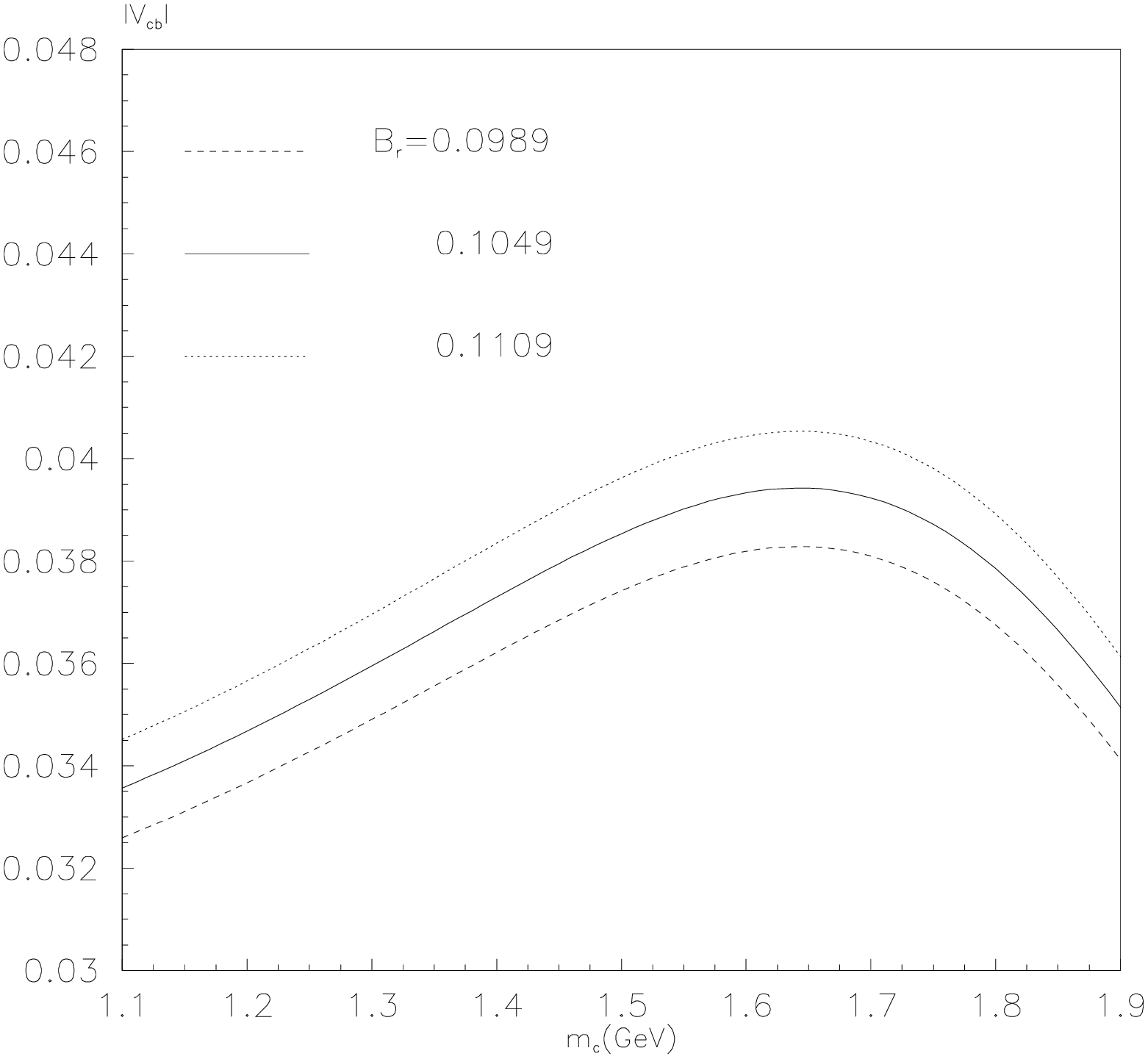} }
\vspace{0cm}
\caption{
  $|V_{cb}|$ extracted from $B\to X_c e\nu$ decay
  at $\mu=2.4\mbox{GeV}$ and $\kappa_1=-0.6\mbox{GeV}^2$.
}
\end{figure}

\end{document}